\documentclass[11pt]{article}

\usepackage{amsmath,amssymb,amsfonts}
\usepackage{graphicx}
\usepackage{cite}
\usepackage{authblk}
\usepackage[colorlinks=true, linkcolor=blue, citecolor=blue, urlcolor=blue]{hyperref}
\usepackage{geometry}
\newtheorem{axiom}{Axiom}
\newtheorem{definition}{Definition}
\newtheorem{theorem}{Theorem}
\newtheorem{corollary}{Corollary}

\title{The Ontic Necessity of the Quantum Wavefunction:\\ Why Epistemic Views Struggle with the Uncertainty Principle}

\author[1]{Bachtiar Rifai}
\author[1]{Muhammad Farchani Rosyid}
\author[1]{Dwi Satya Palupi}

\affil[1]{Cosmology, Astrophysics, Mathematical Physics and Particle Research Group\\
Department of Physics, Universitas Gadjah Mada (UGM), Indonesia\\
\texttt{bachtiar\_rifai@mail.ugm.ac.id}}

\date{}

\begin{document}

\maketitle

\begin{abstract}
The ontological status of the quantum wavefunction remains one of the most debated questions in quantum theory. While epistemic interpretations regard the wavefunction as a reflection of our knowledge or beliefs, ontic interpretations treat it as a real physical object. In this paper, we argue that epistemic approaches struggle to explain the universality and precision of the uncertainty principle, a core feature of quantum mechanics. By contrast, treating the wavefunction as ontic allows a consistent and natural derivation of quantum uncertainty from the mathematical structure of Hilbert space. We examine key interpretations on both sides and highlight why the epistemic view falls short in addressing constraints that appear to be intrinsic to nature.
\end{abstract}

\section{Introduction}

The nature of the quantum wavefunction, $\psi$, lies at the heart of foundational questions in quantum mechanics. Is it a physical entity or merely a mathematical tool for calculating probabilities? This debate divides interpretations of quantum theory into two broad camps: those that treat $\psi$ as \textit{ontic}—a real feature of the world—and those that see it as \textit{epistemic}, representing an agent’s knowledge or beliefs about a system.

Among the most historically influential frameworks is the Copenhagen interpretation, often associated with Niels Bohr and Werner Heisenberg. This view posits that the wavefunction encapsulates all that can be known about a system, and that physical properties become definite only upon measurement \cite{bohr1928, heisenberg1927}. The Copenhagen interpretation thus straddles the ontic–epistemic divide, treating $\psi$ as real in a predictive sense but denying an independent reality prior to observation. While pragmatic, this stance raises philosophical tensions, especially when interpreting quantum phenomena that occur in isolation, without observers.

Historically, the tension between these views dates back to the 1920s. Schrödinger's original wave mechanics~\cite{Schrodinger1926} posited a real wave, but Max Born swiftly recast $|\psi|^2$ as a statistical probability density~\cite{Born1926}. The Copenhagen school, including Bohr and Heisenberg, adopted an instrumentalist stance, treating $\psi$ as a predictor of measurement outcomes with no independent reality. Dissenting voices—most notably Einstein—objected to the purely statistical character of the theory, seeking a more complete description akin to hidden–variable programs. Louis de~Broglie~\cite{DeBroglie1927} and later David Bohm~\cite{bohm1952} proposed pilot‐wave models in which $\psi$ guides point‐particle trajectories, restoring determinism at the cost of manifest nonlocality.

In recent years, theoretical advances such as the Pusey–Barrett–Rudolph (PBR) theorem~\cite{pbr2012} and precision experiments have placed stringent constraints on epistemic reconstructions, showing that any purely epistemic wavefunction must satisfy implausible assumptions (e.g., preparation independence) to avoid contradiction with quantum predictions. These developments have shifted philosophical consensus toward ontic interpretations, though the debate endures—especially when one explores retrocausal, superdeterministic, or epistemically agnostic frameworks.

Our aim in this paper is to show why epistemic interpretations face difficulties in explaining quantum uncertainty as a fundamental, observer-independent feature of nature. We argue that these challenges strengthen the case for an ontic interpretation, in which the wavefunction describes something physically real.

\section{The Epistemic View}

Several modern interpretations present the wavefunction as a state of knowledge rather than a physical object. QBism, for instance, embraces a Bayesian approach: probabilities—and by extension, $\psi$—are personal judgments made by agents \cite{fuchs2014}. In this view, measurement outcomes don't reveal preexisting properties but update beliefs based on experience.

Other approaches, such as Rovelli's relational quantum mechanics \cite{rovelli1996}, argue that physical properties only exist relative to other systems. The wavefunction then encodes the relations between observers and systems, not an absolute state. Similarly, the statistical interpretation, championed by Ballentine \cite{ballentine1970}, maintains that $\psi$ applies to ensembles of systems and not individual particles.

Despite their philosophical diversity, these views share a common stance: the wavefunction is not an element of reality. Instead, it is a tool for reasoning under uncertainty. But here lies the tension.

The Heisenberg uncertainty principle imposes a universal lower bound:
\begin{equation}
  \sigma_x \sigma_p \geq \frac{\hbar}{2}
\end{equation}
This inequality applies even to single particles, regardless of whether a measurement occurs. If $\psi$ merely encodes knowledge, then what exactly is uncertain in a system that has not been observed? Why can’t more precise information simply reduce the uncertainty?

In classical Bayesian reasoning, uncertainty can always be refined with better data. But in quantum mechanics, no amount of information can beat the uncertainty bound. This suggests that quantum uncertainty is not about ignorance—it is built into the structure of the theory itself.

Epistemic interpretations try to account for this by redefining measurement as belief revision or focusing on statistical ensembles. However, such moves seem to dodge the deeper question: why does the universe obey a precise and observer-independent constraint, if $\psi$ is only epistemic?

\section{The Ontic View}

Ontic interpretations offer a more direct answer. If the wavefunction represents a real physical state, then the uncertainty principle follows naturally from the geometry of Hilbert space.

Quantum observables such as position and momentum are represented by noncommuting operators. Their incompatibility leads to uncertainty as a mathematical fact, expressed via the Robertson–Schrödinger inequality \cite{robertson1930}. This relation is not a statement about knowledge, but a property of how physical quantities are encoded in $\psi$.

For example, position and momentum wavefunctions are Fourier transforms of each other. Narrowing one necessarily broadens the other. This trade-off emerges from the wavefunction’s structure, not from our ignorance.

Interpretations like the de Broglie–Bohm theory treat the wavefunction as a guiding field that determines a particle’s motion \cite{bohm1952}. The particle has a definite position, but its dynamics are fully determined by $\psi$, which is taken as physically real. In the many-worlds interpretation, the universal wavefunction never collapses, and measurement merely reveals the branch we occupy \cite{everett1957}. Again, $\psi$ is real and its evolution is unitary.

The results of Sharoglazova \emph{et al.} \cite{Sharoglazova2025} expose a discrepancy between inherently stochastic quantum predictions and the deterministic trajectories posited by Bohmian mechanics.  In addition to challenging the guiding equation, these findings offer broader philosophical lessons about the ontological status of the wavefunction. Sharoglazova \emph{et al.} performed a precise measurement of the energy–speed relationship in a coupled-waveguide experiment, demonstrating that particles with increasingly negative local kinetic energy traverse potential steps at higher effective velocities \cite{Sharoglazova2025}.  Crucially, these observations diverge from Bohmian forecasts, which would predict vanishing speed under such conditions and therefore infinite dwell times.  This empirical refutation lends credence to the orthodox interpretation, wherein the wavefunction directly governs measurement statistics without recourse to hidden trajectories.  Moreover, the authors note that no modified guiding equation can reconcile both the dwell-time data and the standard quantum mechanical current density simultaneously, reinforcing the wavefunction's primacy as the descriptor of physical state rather than mere epistemic bookkeeping.

Perhaps most striking is the Pusey–Barrett–Rudolph (PBR) theorem \cite{pbr2012}, which shows that under certain assumptions, an epistemic view of the wavefunction leads to contradictions with quantum predictions. If two distinct wavefunctions can correspond to the same physical state, then some outcomes must violate the Born rule. But experiments show no such violations.

In addition, the Colbeck–Renner theorem \cite{colbeck2012} argues that the wavefunction must be in one-to-one correspondence with the underlying reality, assuming free choice and no superdeterminism. Here, superdeterminism—the idea that measurement settings and system properties are correlated due to a common causal past—is sometimes invoked to escape the theorem’s implications \cite{hossenfelder2020}. However, such assumptions challenge the very foundation of scientific reasoning, as they undermine the notion of independent experimental control.

A paradigmatic example of how theoretical simplicity can precipitate profound physical insights is provided by the Dirac equation.  By seeking a linear relativistic wave equation of the form
\begin{equation}
(i\gamma^\mu \partial_\mu - m )\psi = 0,
\end{equation}
Dirac discovered that consistency with the relativistic energy–momentum relation
\begin{equation}
E^2 = p^2c^2 + m^2c^4
\end{equation}
necessitates solutions with both positive and negative energy eigenvalues, $E=\pm\sqrt{p^2c^2+m^2c^4}$.  While the positive branch was naturally identified with electrons, the negative branch initially presented a paradox. Dirac resolved this by postulating a filled "sea" of negative-energy states, whose holes manifest as positrons, the electron's antiparticle. This remarkable prediction of antimatter, confirmed by Anderson's discovery of the positron in 1932, underscores that entities of fundamental physical reality may simply emerge from the mathematical structure itself: what begins as algebraic necessity becomes empirical truth.  Thus, the Dirac equation offers a compelling case in which "reality can emerge from simple mathematics."

\section{Concrete Examples}

Quantum experiments provide concrete insight into the ontological status of the wavefunction. We examine several key examples demonstrating that the wavefunction $\psi$ must correspond to real physical structure, not merely knowledge or belief.

A canonical illustration is the double-slit experiment with single particles. When a quantum particle such as an electron or photon passes through a pair of slits, and no which-path information is obtained, an interference pattern appears on the detection screen. The wavefunction evolves according to the Schrödinger equation and is coherently split by the slits. If $\psi_A(x)$ and $\psi_B(x)$ represent the amplitudes associated with slits $A$ and $B$ at a screen position $x$, then the total wavefunction is
\begin{equation}
    \psi(x) = \psi_A(x) + \psi_B(x).
\end{equation}
The detection probability at position $x$ is given by the Born rule:
\begin{equation}
    P(x) = |\psi(x)|^2 = |\psi_A(x)|^2 + |\psi_B(x)|^2 + 2 \text{Re}[\psi_A^*(x)\psi_B(x)],
\end{equation}
where the interference term $2 \text{Re}[\psi_A^*(x)\psi_B(x)]$ is nonzero due to coherent superposition. This pattern emerges even when only one particle passes through the apparatus at a time, demonstrating that interference is a property of the wavefunction itself—not of ensembles or subjective uncertainty. The persistence of interference under conditions where no observer possesses which-path knowledge supports an ontic reading of $\psi$.

Even stronger implications arise from Bell test experiments. Consider two entangled qubits prepared in the singlet state:
\begin{equation}
    |\psi\rangle = \frac{1}{\sqrt{2}}(|0\rangle_A|1\rangle_B - |1\rangle_A|0\rangle_B).
\end{equation}
Let Alice and Bob each choose between two measurement settings, $A, A'$ for Alice and $B, B'$ for Bob, yielding binary outcomes $\pm1$. Assuming realism, locality, and measurement independence, Bell derived the inequality:
\begin{equation}
    |E(A,B) + E(A,B') + E(A',B) - E(A',B')| \leq 2,
\end{equation}
where $E(X,Y)$ denotes the expectation value of the product of outcomes from measurements $X$ and $Y$. Quantum theory predicts that for certain choices of observables, this bound is violated, with the maximal violation known as the Tsirelson bound:
\begin{equation}
    |E(A,B) + E(A,B') + E(A',B) - E(A',B')| = 2\sqrt{2}.
\end{equation}
This violation has been confirmed in loophole-free experiments, such as in \cite{hensen2015}, and cannot be explained by any local hidden variable theory. An epistemic view of $\psi$ must resort to highly contrived mechanisms such as superdeterminism to preserve locality, implying that hidden variables conspire with measurement settings to simulate quantum correlations. The ontic view straightforwardly accounts for these nonlocal correlations as physical consequences of an entangled $\psi$.

Weak measurement techniques provide further evidence for the reality of the wavefunction. Given a preselected state $|\psi\rangle$ and postselected state $|\phi\rangle$, the weak value of an observable $A$ is given by:
\begin{equation}
    \langle A \rangle_w = \frac{\langle \phi | A | \psi \rangle}{\langle \phi | \psi \rangle}.
\end{equation}
Weak values can lie outside the spectrum of $A$ and have been measured experimentally using interaction-weak regimes that minimally disturb the system. Such measurements reconstruct intermediate values in quantum trajectories and have enabled weak tomography of quantum systems \cite{aharonov1988}. The statistical reliability and reproducibility of weak values suggest that $\psi$ is more than a belief update mechanism; it encodes objective properties that guide system evolution even between measurements.

A deeper structural argument for the reality of $\psi$ arises from the geometry of Hilbert space. In standard quantum theory, the state of a system is represented by a unit vector $|\psi\rangle$ in a complex Hilbert space $\mathcal{H}$. Observables are Hermitian operators $A$ on $\mathcal{H}$, and the Born rule assigns outcome probabilities via:
\begin{equation}
    P(a_i) = \langle \psi | \Pi_{a_i} | \psi \rangle,
\end{equation}
where $\Pi_{a_i}$ is the projector onto the eigenspace of $A$ with eigenvalue $a_i$. More generally, quantum measurements are described by positive operator-valued measures (POVMs), which associate each possible outcome $a_i$ with a positive semi-definite operator $E_i$ satisfying $\sum_i E_i = \mathbb{I}$. The probability of outcome $i$ is then:
\begin{equation}
    P(i) = \langle \psi | E_i | \psi \rangle.
\end{equation}
This formalism shows that outcome statistics are determined by the inner product structure of $\mathcal{H}$ and the full amplitude distribution encoded in $\psi$. An epistemic view would need to reinterpret all of this geometry in terms of belief spaces and subjective assignments, a task that appears less parsimonious and less consistent with the formal coherence of the theory.

Finally, decoherence provides a dynamical explanation for the emergence of classicality without invoking wavefunction collapse. When a quantum system interacts with an environment, the reduced density matrix becomes approximately diagonal in a pointer basis due to entanglement with the environment:
\begin{equation}
    \rho_S(t) = \text{Tr}_E [U(t)(\rho_S(0) \otimes \rho_E(0))U^\dagger(t)] \approx \sum_i p_i |i\rangle\langle i|.
\end{equation}
This process suppresses interference and explains the apparent transition from quantum to classical behavior in measurement-like contexts. Crucially, the total wavefunction evolves unitarily, and the appearance of classical outcomes is due to tracing out environmental degrees of freedom. This supports the ontic status of $\psi$ as a physically evolving entity, not a collapsing belief state.

Together, these experiments and theoretical constructions converge on the conclusion that the wavefunction is not a mere tool for encoding subjective knowledge. Its mathematical structure, empirical predictions, and explanatory power indicate that $\psi$ corresponds to something real—albeit nonclassical—in the ontology of quantum systems.

\section{Incompatibility of Epistemic Views with Uncertainty Principle}

This section presents a formal argument that epistemic interpretations of the wavefunction are fundamentally incompatible with the Heisenberg uncertainty principle. The reasoning is developed using axioms grounded in quantum theory, Hilbert space geometry, and classical epistemic frameworks.

\begin{axiom}[Hilbert Space Formalism]
Every quantum system is described by a normalized state vector $\psi \in \mathcal{H}$, where $\mathcal{H}$ is a complex Hilbert space. Observables correspond to Hermitian operators on $\mathcal{H}$.
\end{axiom}

This is a foundational postulate of quantum theory. The spectral theorem guarantees that Hermitian operators correspond to measurable quantities, and the Born rule determines outcome probabilities via inner products in $\mathcal{H}$.

\begin{axiom}[Robertson–Schrödinger Uncertainty Principle]
For any state $\psi \in \mathcal{H}$ and two non-commuting observables $A$ and $B$, the following holds:
\begin{equation}
\sigma_A(\psi) \cdot \sigma_B(\psi) \geq \frac{1}{2} \left| \langle [A, B] \rangle_\psi \right|.
\end{equation}
For $\hat{x}$ and $\hat{p}$:
\begin{equation}
\sigma_x \cdot \sigma_p \geq \frac{\hbar}{2}.
\end{equation}
\end{axiom}

This follows from the Cauchy–Schwarz inequality applied in Hilbert space to the vectors $A \psi$ and $B \psi$, along with the non-commutativity of $A$ and $B$. For position and momentum, $[\hat{x}, \hat{p}] = i\hbar$ directly yields the lower bound.

\begin{axiom}[Epistemic Interpretation]
In epistemic models, the wavefunction $\psi$ encodes incomplete knowledge of an underlying ontic state $\lambda \in \Lambda$. There exists a probability distribution $\mu_\psi(\lambda)$ such that:
\begin{equation}
\psi \sim \mu_\psi(\lambda).
\end{equation}
\end{axiom}

This formulation captures the essence of epistemic interpretations such as QBism and statistical ensembles, where $\psi$ does not correspond directly to physical reality but represents belief or preparation knowledge over $\Lambda$.

\begin{definition}[Bayesian Refinement]
In classical epistemology, additional data leads to sharper distributions via Bayesian updating:
\begin{equation}
\mu_\psi(\lambda) \rightarrow \mu_{\psi'}(\lambda),
\end{equation}
with
\begin{equation}
\text{Var}^{\mu_{\psi'}}(x) < \text{Var}^{\mu_\psi}(x), \quad \text{Var}^{\mu_{\psi'}}(p) < \text{Var}^{\mu_\psi}(p).
\end{equation}
In the limit of perfect knowledge:
\begin{equation}
\sigma_x, \sigma_p \rightarrow 0 \quad \Rightarrow \quad \sigma_x \cdot \sigma_p \rightarrow 0.
\end{equation}
\end{definition}

\begin{theorem}[Contradiction with Quantum Theory]
The Bayesian refinement permitted by the epistemic interpretation allows $\sigma_x \cdot \sigma_p \rightarrow 0$, which contradicts the quantum lower bound:
\begin{equation}
\sigma_x \cdot \sigma_p \geq \frac{\hbar}{2} > 0.
\end{equation}
\end{theorem}

Under the epistemic view, variance reduction is unconstrained in principle, because $\lambda$ is assumed to exist independently of $\psi$, and more accurate knowledge refines $\mu_\psi(\lambda)$. However, in quantum theory, the uncertainty principle arises from the structural properties of the wavefunction in $\mathcal{H}$ and holds universally, even in the absence of measurement. Therefore, if $\psi$ is merely epistemic, the epistemic model must violate the Robertson–Schrödinger relation in the limit, leading to a contradiction.

\begin{corollary}[Ontic Necessity of the Wavefunction]
Any interpretation in which:
\begin{enumerate}
    \item the wavefunction is not uniquely determined by the ontic state, and
    \item uncertainty is attributed solely to incomplete knowledge,
\end{enumerate}
is incompatible with the quantum uncertainty principle. Therefore, $\psi$ must encode real, irreducible physical properties—that is, it must be ontic.
\end{corollary}

\section{Conclusion}

The Heisenberg uncertainty principle is not just a practical limit on what we can know. It is a structural feature of quantum theory, rooted in the noncommutative algebra of observables and the geometry of Hilbert space.

Epistemic interpretations, despite offering elegant philosophical frameworks, struggle to explain why such constraints apply universally—even in unmeasured systems. They must treat the uncertainty principle as a rule about beliefs, yet its precision and universality suggest it is something deeper. We conclude that the wavefunction $\psi$ cannot merely reflect incomplete knowledge about an underlying reality. Rather, it must encode irreducible physical features of the system—i.e., it must be ontic.

Ontic interpretations, by taking the wavefunction as real, provide a straightforward explanation: uncertainty arises from the very structure of physical reality. The success of quantum theory, the predictions of no-go theorems, and the results of key experiments all point in this direction.

\section*{Acknowledgments and Future Directions}

We acknowledge that the present argument, while formally demonstrating a tension between epistemic interpretations and the uncertainty principle, does not encompass all possible variants of epistemic models. In particular, interpretations incorporating retrocausality, contextuality, or epistemic–ontic hybrids may warrant further examination. Additionally, our analysis presumes the standard Hilbert space framework and does not engage with alternative formulations such as generalized probabilistic theories (GPTs) or quantum causal models.

Future research may explore the following directions:
\begin{itemize}
    \item Examining whether non-classical probability structures or epistemic toy models can reconcile epistemic interpretations with the empirical and structural features of quantum theory.
    \item Investigating the interplay between contextuality, nonlocality, and uncertainty to more precisely delineate the limits of epistemic reconstructions.
    \item Proposing experimental tests that can operationally distinguish ontic from epistemic accounts of quantum uncertainty.
    \item Extending the argument to open systems and decoherence frameworks to evaluate the resilience of epistemic interpretations in dynamically evolving environments.
\end{itemize}

We thank the Department of Physics, Faculty of Mathematics and Natural Sciences, Universitas Gadjah Mada (UGM), for their support. We are also grateful to colleagues in the quantum foundations community for discussions that informed and sharpened the core ideas of this work.

\end{document}